\documentclass[aps,showpacs,preprintnumbers,amsmath,amssymb]{revtex4}
\usepackage{dcolumn}
\usepackage[dvips]{epsfig}
\usepackage{graphicx}
\usepackage{amssymb}
\usepackage{color}
\usepackage{enumerate}
\usepackage{subfigure}

\begin{document}
\baselineskip=0.8 cm
\title{\bf Dynamical evolution of electromagnetic perturbation in the background of a scalar hairy black
hole in Horndeski theory }

\author{Qiong Fang$^{1}$,  Songbai Chen$^{1,2,3,4}$\footnote{Corresponding author: csb3752@hunnu.edu.cn}, Jiliang
Jing$^{1,2,3,4}$\footnote{jljing@hunnu.edu.cn}
}
\affiliation{$ ^1$ Department of Physics, Hunan
Normal University,  Changsha, Hunan 410081, People's Republic of
China \\ $ ^2$Key Laboratory of Low Dimensional Quantum Structures \\
and Quantum Control of Ministry of Education, Hunan Normal
University, Changsha, Hunan 410081, People's Republic of China\\
$ ^3$Synergetic Innovation Center for Quantum Effects and Applications,
Hunan Normal University, Changsha, Hunan 410081, People's Republic
of China\\
$ ^4$Center for Gravitation and Cosmology, College of Physical Science and Technology,
Yangzhou University, Yangzhou 225009, People's Republic
of China}

\begin{abstract}
\baselineskip=0.6 cm
\begin{center}
{\bf Abstract}
\end{center}

We have investigated dynamical evolution of electromagnetic perturbation in a scalar hairy black hole spacetime, which belongs to solutions in Horndeski theory with a logarithmic cubic term. Our results show that the parameter $\alpha$ affects the existence of event horizon and modifies the asymptotical structure of spacetime at spatial infinity, which imprints on the quasinormal frequency of electromagnetic perturbation. Moreover, we find that the late-time tail of electromagnetic perturbation
in this background depends also on the parameter $\alpha$ due to the existence of solid angle deficit. The presence of the parameter $\alpha$ makes the perturbation field decay more rapidly. These imply that the spacetime properties arising from the logarithmic cubic term in the action play important roles in the dynamical evolutions of the electromagnetic perturbation in the background of a scalar hairy black hole.

\end{abstract}

 \maketitle
\newpage
\section{Introduction}

The importance of studying dynamical evolution of  external perturbations
around a black hole is beyond any reasonable doubt because these dynamical evolutions are deeply connected  with many important issues including the stability and identification of the black holes. In a black hole spacetime, the dynamical evolution for any kind of perturbations consists of three stages ( for a review, see \cite{evolu1,evolu2,evolu3,evolu4} ). The first stage is the initial outburst of the waves which completely depends on only the initial form of perturbation field. The second one involves the damped oscillation which is described by quasinormal
mode with complex frequency. The real part of quasinormal frequency  denotes
the real frequency of the oscillation and the imaginary
part represents the damping. The last one is the power-law tail  of
the perturbation field caused by backscattering of the gravitational field at very late time.
It is certain that the quasinormal mode
carries characteristic fingerprint of a black hole and it could offer a direct way to identify the black hole existence in Universe. The detection of quasinormal modes is expected to be realized through high sensitive detectors of gravitational wave in the near future. Apart from the astrophysical interest, the quasinormal mode has also been argued as a useful tool to test ground for fundamental physics, such as, the quantum gravity \cite{quan4,quan5,quan6} and the AdS/CFT correspondence \cite{ads7,ads8,ads9}, and also the phase transition of black holes \cite{phst1,phst2,phst3}. At the late-time stage, it is well known that the behaviors of the massless
perturbations fields for a fixed $r$ are dominated by the factor $t^{-(2l+3)}$ in the usual four-dimensional spacetimes \cite{latt1,latt2,latt3}. For the massive perturbations, the late-time tails are dominated by the oscillations with inverse power-law form $t^{-(l+3/2)}\sin\mu t$ in the intermediate late time \cite{lattm1,lattm2,lattm3}. However, in the spacetimes with a deficit solid angle ( caused by global monopole \cite{lattdf1}, cosmic string \cite{lattdf2} or tensional brane \cite{lattdf3} ), it is found that the late-time tails of perturbations also depend on the symmetry breaking
scale. Moreover, dynamical evolutions of perturbation fields in the background of a compact object without horizon (such as, neutron stars \cite{qnns1,qnns2,qnns3}, wormholes \cite{qnwh0,qnwh1,qnwh2,qnwh3,qnwh4}) have been investigated recently, which indicates that there exist gravitational echoes. Such kinds of gravitational echoes do not appear in the black hole background. This can be attributed to the difference of the boundary conditions of perturbation field in these two kinds of background spacetimes.

Einstein's general relativity is considered probably the most beautiful theory of gravity at present. However, there are a lot of investigations focusing on the modification to Einstein's theory of gravity since modifying general relativity is one of promising ways to explain the accelerating expansion of the current Universe observed through astronomical experiments. Horndeski theory is such kind of alternative theories of gravity \cite{hornth}. In Horndeski theory, the action contains higher derivatives of both metric $g_{\mu\nu}$ and the scalar field $\phi$, but the equations of motion are at most the second order. Moreover, even though there are higher derivative terms, there exists no Ostrogradsky instability. These means that fields in Horndeski theory have good dynamical behaviors, which triggers the extensive applications of Horndeski theory in black hole physics and cosmology, particularly in the construction of cosmological models of inflation and dark energy (for a review, see  \cite{hornth1,hornth2} ). However, the free parameters of Horndeski theory, especially the modifications from the Lagrangians $L_4$ and $L_5$, are strongly constrained by the direct measurement of the speed of gravitational waves following GW170817 \cite{htcons1,htcons2,htcons3,htcons4,htcons5}. This means that the range of possible modifications from Horndeski theory has been reduced drastically. Recently, Tattersall \textit{et al} \cite{tatt} have looked at the allowed space of the actions in Horndeski theory to see which of them will lead to distinctive signatures around
black holes. After adding a logarithmic cubic term \cite{htact1,tatta1} in the action of Einstein's general relativity with a scalar field, they obtain a static scalar hairy black hole solution with three free parameters. This solution is not asymptotically flat due to the existence of a deficit solid angle, which could imprint in the dynamical evolution of matter field  propagating in the spacetime.
In this paper, we will study dynamical evolution of electromagnetic perturbation in this scalar hairy black hole spacetime.

The paper is organized as follows. In Sec. II, we review briefly the scalar hairy black hole solution obtained by Tattersall \textit{et al} \cite{tatt}, and then analyse its spacetime structure and the corresponding thermodynamical properties. In Sec.III, we study quasinormal modes and late-time tail of electromagnetic perturbation in this background and probe effects of the spacetime parameters on the dynamical evolution of electromagnetic perturbation.
Finally, we present a summary.

\section{a scalar hairy black hole in Horndeski theory}

Let us now to introduce briefly the black hole solution with scalar hair in the Horndeski theory.
The most general action for the Horndeski theory with electromagnetic field can be expressed as \cite{hornth}
\begin{eqnarray}\label{acts}
S=\int d^4x\sqrt{-g}\sum_{n=2}^5 \bigg[L_n-\frac{1}{4}F_{\mu\nu}F^{\mu\nu}\bigg],
\end{eqnarray}
where the Lagrangians $L_n$ are given by
\begin{eqnarray}
L_2&=&G_2(\phi,X),\\
L_3&=&-G_3(\phi,X)\Box \phi,\\
L_4&=&G_4(\phi,X)R+G_{4,X}(\phi,X)((\Box\phi)^2-\phi^{\mu\nu}\phi_{\mu\nu} ),\\
L_5&=&G_5(\phi,X)G_{\mu\nu}\phi^{\mu\nu}-\frac{1}{6}G_{5,X}(\phi,X)((\Box\phi)^3-3\box\phi \phi^{\mu\nu}\phi_{\mu\nu} +2 \phi_{\mu\nu}\phi^{\mu\sigma}\phi^{\nu}_{\sigma}),
\end{eqnarray}
respectively. Here $\phi$ is the scalar field with kinetic term $X=-\phi_\mu\phi^\mu/2$ where $\phi_\mu=\nabla_\mu\phi$. The quantity $\phi_{\mu\nu}=\nabla_\nu\nabla_\mu\phi$, and the tensor $G_{\mu\nu}$ is the well-known Einstein tensor with the form $G_{\mu\nu}=R_{\mu\nu}-\frac{1}{2}R\,g_{\mu\nu}$. The $G_i$ are arbitrary functions of $\phi$ and $X$, and $G_{i,X}$ correspond to the derivatives of $G_i$ with respect to $X$. $F_{\mu\nu}$ is
the electromagnetic tensor, which is related to the electromagnetic
vector potential $A_{\mu}$ by $F_{\mu\nu}=A_{\nu;\mu}-A_{\mu;\nu}$.
In order to obtain a black hole solution, one must solve the equations of motion for the gravitational field and scalar field which can be obtained by varying the action (\ref{acts}). In general, it is difficult  to get an analytical solution in the Horndeski theory (\ref{acts})  even if the equations of motion are at most second order. Recently, Tattersall \textit{et al} \cite{tatt} obtain a four-dimensional static and spherical
symmetric solution of black hole with scalar hair by choosing the action containing an logarithmic cubic term, i.e.,
\begin{eqnarray}
G_2(\phi,X)=X,\;\;\;\;\;\;\;\;\;\;\;\; G_3(\phi,X)=-\alpha M_P\log(-X),\;\;\;\;\;\;\;\;\;\;\;\;G_4(\phi,X)=M^2_P/2, \;\;\;\;\;\;\;\;\;\;\;\; G_5(\phi,X)=0,
\label{actbh1}
\end{eqnarray}
where $\alpha$ is an arbitrary dimensionless
constant and $M_P$ is the usual Plank mass. The choice of above actions is based on a consideration that the modifications from the actions $L_4$ and $L_5$ are strongly constrained by the direct measurement of the speed of gravitational waves following GW170817 \cite{htcons1,htcons2,htcons3,htcons4,htcons5}. Actually, the action (\ref{actbh1}) describes the modification to Einstein's general relativity with a scalar field by a logarithmic cubic term, which ensures the existence of the black hole solution with a scalar hairy \cite{tatt}. It is found that in a standard hydrodynamical description for  scalar
field theories this logarithmic cubic form of the $G_3(\phi,X)$ implies the
diffusivity (i.e., a coefficient between the spatial gradient of the chemical potential and energy flow ) is just to be constant \cite{htact1}. The metric of this hairy black hole solution has a form
\begin{eqnarray}
ds^2=-\left(1-\frac{2M}{r}+\frac{c}{r^{4+\frac{1}{\alpha^2}}}\right)dt^2+\left(1-\frac{2M}{r}+\frac{c}{r^{4+\frac{1}{\alpha^2}}}\right)^{-1}dr^2
+\frac{r^2}{1+4\alpha^2}(d\theta^2+\sin^2\theta d\varphi^2),\label{m1}
\end{eqnarray}
where the parameter $M$ is an integration constant related to the mass of black hole as in the Schwarzschild case, and the constant $c$ is related to the scalar field $\phi$. It is obvious that the spacetime (\ref{m1}) is not asymptotically flat since there is a deficit solid angle $ 4\pi(1+4\alpha^2)^{-1}$, which is similar to that in the global monopole spacetime. Moreover,  the deficit solid angle decreases with the parameter $\alpha$. When $\alpha$ tends to zero, the metric (\ref{m1}) approaches to that of Schwarzschild spacetime.

\begin{figure}
\center
\includegraphics[width=6cm]{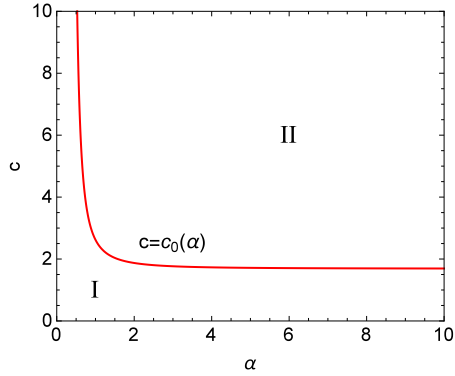}
\caption{The dependence of the existence of horizon on the parameter $\alpha$ and $c$. The regions $I$ and $II$ separated by curves $c=c_0$ are corresponded to the cases of the spacetime (\ref{m1}) with horizons and no any horizon, respectively. Here, we set $M=1$.}
\label{s5}
\end{figure}

The position of event horizon of the black hole lies in
\begin{eqnarray}\label{shjie}
r^{4+\frac{1}{\alpha^2}}-2Mr^{3+\frac{1}{\alpha^2}}+c=0.
\end{eqnarray}
Solving this equation, we find that there exist a
critical value for the existence of horizon in the spacetime (\ref{m1})
\begin{eqnarray}
c_0=\frac{\alpha^2}{1+3\alpha^2}\bigg[\frac{2M(1
+3\alpha^2)}{1+4\alpha^2}\bigg]^{4+\frac{1}{\alpha^2}},\label{evodsch}
\end{eqnarray}
which decreases with the parameter $\alpha$. As shown in Fig.(\ref{s5}),  the critical curve $c_0(\alpha)$ in the plane ($\alpha$, $c$) splits the total plane into two regions $I$ and $II$. In the region $I$ with $c<c_0$, we find that there exist two horizons for the spacetime (\ref{m1}), which is similar to that in the Reissner Nordstr\"{o}m black hole spacetime.  As the parameters lie in the region $II$ (i.e., $c>c_0$), we can obtain that there is no horizon and the metric describes
the geometry of a naked singularity.
If $c$ equals to the critical value $c_0$, two black hole horizons degenerate and then the black hole is extremal.  As $\alpha$ tends to zero, from Eqs.(\ref{shjie}) and (\ref{evodsch}), one can find that the radius of event horizon $r_h$ approaches to $2M$ and the vaule of $c_0$ tends to infinity, which means that the horizon exists always and there is only a single horizon in this limit.
Now, let us further discuss  thermodynamical property of the
black hole (\ref{m1}). Following Ref. \cite{energy1}, we find that the energy $E$ of the black hole can be written as
\begin{eqnarray}
E=\frac{M}{1+4\alpha^2},
\end{eqnarray}
which depends on the parameter $\alpha$.
The entropy $S$, energy $E$ and Hawking temperature $T$ of the black hole (\ref{m1})
can be described by
\begin{eqnarray}
S&=&\frac{A_h}{4}=\frac{\pi r^2_h}{1+4\alpha^2},\\
E&=&\frac{1}{2}\sqrt{\frac{S}{1+4\alpha^2}}\bigg[\frac{1}{\sqrt{\pi}}+\frac{c\,\pi^{\frac{3\alpha^2+1}{2\alpha^2}}}{
[(1+4\alpha^2)S]^{\frac{4\alpha^2+1}{2\alpha^2}}}\bigg],\\
T&=&\bigg(\frac{\partial E}{\partial S}\bigg)_c=\frac{1}{4\pi r_h}\bigg[1-\frac{c\,(1+3\alpha^2)}{\alpha^2r^{(\frac{4\alpha^2+1}{\alpha^2})}_h}\bigg].
\end{eqnarray}
Treating the constant $c$ as a
variable, one can the generalized force
\begin{eqnarray}
\Theta_c=\bigg(\frac{\partial E}{\partial c}\bigg)_S=\frac{\pi^{\frac{3\alpha^2+1}{2\alpha^2}}}{
(1+4\alpha^2)^{\frac{5\alpha^2+1}{2\alpha^2}}S^{\frac{3\alpha^2+1}{2\alpha^2}}}=\frac{1}{2(4\alpha^2+1)}
r^{-(\frac{3\alpha^2+1}{\alpha^2})}_h.
\end{eqnarray}
And then we find that the first law takes the form
\begin{eqnarray}\label{TS1}
E=2TS+\frac{1+4\alpha^2}{\alpha^2}\Theta_c c.
\end{eqnarray}
Obviously, in the case of a scalar hairy black hole (\ref{m1}), the form of the
first law (\ref{TS1}) depend on both the parameters $\alpha$ and $c$.
The differential form of the first law
 becomes
\begin{eqnarray}\label{TS2}
dE=TdS+\Theta_c dc.
\end{eqnarray}
which is explicitly independent of the parameter $\alpha$.
When $\alpha$ tends to zero, one can find that the quantities
\begin{eqnarray}
(E-2TS)\bigg|_{\alpha\rightarrow 0}=\frac{2M-r_h}{2}\bigg[\frac{\ln\frac{c}{2M-r_h}}{\ln r_h}-3\bigg]\bigg|_{r_h\rightarrow 2M}=0,
\end{eqnarray}
which means that  the forms of the first law (\ref{TS1}) and (\ref{TS2}) approach to those of Schwarzschild black hole in this limit.

\section{Quasinormal modes and late-time tail of electromagnetic perturbation in the background of a scalar hairy black hole }

In this section, we focus only on the case where the horizon exists and then study the dynamical evolution of the electromagnetic perturbations in this background.
For the scalar hairy black hole spacetime (\ref{m1}), one can expand electromagnetic vector $A_{\mu}$ in spherical harmonics \cite{emagnet}
\begin{eqnarray}
A_{\mu}= \sum_{l,m}\left(\left[\begin{array}{ccc}
 &0&\\
 &0&\\
 &\frac{a^{lm}(t,r)}{\sin\theta}\partial_{\phi}Y_{lm}&\\
 &-a^{lm}(t,r)\sin\theta\partial_{\theta}Y_{lm}&
\end{array}\right]+\left[\begin{array}{cccc}
 &j^{lm}(t,r)Y_{lm}&\\
 &h^{lm}(t,r)Y_{lm}&\\
 &k^{lm}(t,r)\partial_{\theta}Y_{lm}&\\
 &k^{lm}(t,r)\partial_{\phi}Y_{lm}&
\end{array}\right]\right),\label{Au}
\end{eqnarray}
where $l$ and $m$ are the angular quantum
number and the azimuthal number, respectively. The first term in the right side has parity $(-1)^{l+1}$ and the second term has parity $(-1)^{l}$.
Adopting the following form
\begin{eqnarray}
a^{lm}(t,r)&=&a^{lm}(r)e^{-i\omega t},~~~~~h^{lm}(t,r)=h^{lm}(r)e^{-i\omega t},\\
j^{lm}(t,r)&=&j^{lm}(r)e^{-i\omega t},~~~~~k^{lm}(t,r)=k^{lm}(r)e^{-i\omega t},
\end{eqnarray}
and then inserting the above expansion (\ref{Au})
into the usual Maxwell equation,
\begin{eqnarray}
\nabla_{\mu}F^{\mu\nu}=0.\label{WE}
\end{eqnarray}
one can obtain three independent coupled differential equations. Eliminating $k^{lm}(r)$, we can get a single second order differential equation for the electromagnetic perturbation
\begin{eqnarray}
\frac{d^2\Psi(r)}{dr^2_*}+[\omega^2-V(r)]\Psi(r)=0,\label{radial}
\end{eqnarray}
where $r_{*}$ is the tortoise coordinate defined as $dr_*=\frac{
dr}{1-2Mr^{-1}+cr^{-(4+1/\alpha^2)}}$. The wave function $\Psi(r)$ is a linear combination of the functions $j^{lm}(r)$, $h^{lm}(r)$, and $a^{lm}(r)$ in the expansion (\ref{Au}).
The form of $\Psi(r)$ depends on the
parity of the perturbation. For the odd parity $(-1)^{l+1}$, $\Psi(r)$ has the form
\begin{eqnarray}
\Psi(r)=a^{lm},
\end{eqnarray}
but for the even parity $(-1)^{l}$, it has
\begin{eqnarray}
\Psi(r)=\frac{r^2}{l(l+1)}\bigg(-i\omega
h^{lm}-\frac{dj^{lm}}{dr}\bigg).
\end{eqnarray}
However, the effective
potential $V(r)$ in Eq. (\ref{radial}) is independent of the parity of
the perturbation, which is the same as that of free electromagnetic perturbation in the usual static black hole spacetimes. The effective potential $V(r)$ in Eq. (\ref{radial}) can be written as
\begin{eqnarray}\label{veff}
V(r)=\bigg(1-\frac{2M}{r}+\frac{c}{r^{4+\frac{1}{\alpha^2}}}\bigg)\frac{l(l+1)(1+4\alpha^2)}{r^2}.
\end{eqnarray}
It is obvious that the parameters $\alpha$ and $c$ emerge in the
effective potential, which will affect
the dynamical evolution of the electromagnetic perturbation in the scalar hairy black
hole spacetime (\ref{m1}).

Without loss of generality, we here focus on the fundamental quasinormal frequencies of perturbation field. To obtain the fundamental quasinormal frequencies of the electromagnetic perturbation around the scalar hairy black hole (\ref{m1}), we can adopt the 6th-order WKB approximation \cite{wkb2}. The formula for the complex quasinormal frequencies $\omega$ in this approximation can be expressed as
\begin{eqnarray}
\frac{i(\omega^2-V_0)}{\sqrt{-2V^{''}_0}}-\Lambda_2-\Lambda_3-\Lambda_4-\Lambda_5-\Lambda_6
=n+\frac{1}{2},
\end{eqnarray}
with
\begin{eqnarray}
V^{(s)}_0=\frac{d^sV}{dr^s_*}\bigg|_{\;r_*=r_*(r_{p})},
\end{eqnarray}
$n$ is the overtone number and $r_p$ is the value of polar coordinate $r$ corresponding to the
peak of the effective potential (\ref{veff}). The expressions for $\Lambda_2\sim \Lambda_6$ can be
found in \cite{wkb2,wkb}. With this formula and the effective potential (\ref{veff}), we can get the quasinormal frequencies of electromagnetic field in the scalar hairy black hole spacetime (\ref{m1}).
\begin{table}[ht]
\center
\begin{tabular}[b]{cccc}
 \hline \hline
 \;\;\;\; $c$ \;\;\;\; & \;\;\;\; $\omega\ \ \ (\alpha=0.3)$\;\;\;\;  & \;\;\;\;  $\omega \ \ \ (\alpha=0.4)$\;\;\;\;& \;\;\;\; $\omega \ \ \ (\alpha=0.5)$ \;\;\;\; \\ \hline &&&
\\
2.0&\; 0.29682322-0.09357821i \;&\; 0.32990851-0.09363878i \;&\; 0.36832957-0.09331716i \;
 \\
4.0&0.29677745-0.09358955i&0.32995146-0.09327107i&0.36856935-0.09223555i
 \\
6.0&0.29673168-0.09360091i&0.32998572-0.09291139i&0.36869985-0.09118005i
 \\
8.0&0.29668560-0.09361225i&0.33001190-0.09255971i&0.36872679-0.09016725i
\\
10 &0.29664031-0.09362360i&0.33003039-0.09221610i&0.36866636-0.08921202i
\\
12 &0.29659467-0.09363494i&0.33004187-0.09188048i&0.36854396-0.08832390i
\\
 \hline \hline
 $c$  &$\omega \ \ \ (\alpha=0.55)$ & $\omega \ \ \ (\alpha=0.60)$ & $\omega \ \ \ (\alpha=0.70)$\\ \hline&&&
 \\
0.1&0.38854332-0.09450623i&0.40987271-0.09463973i&0.45445760-0.09485755i
 \\
0.5&0.38866776-0.09420939i&0.41007124-0.09425718i&0.45489493-0.09429323i
 \\
1.0&0.38881444-0.09383431i&0.41030782-0.09376774i&0.45542863-0.09355292i
 \\
2.0&0.38907536-0.09307298i&0.41073483-0.09275470i&0.45642893-0.09195574i
\\
3.0&0.38928844-0.09230130i&0.41108730-0.09170460i&0.45729009-0.09021547i
\\
4.0&0.38944931-0.09152562i&0.41135200-0.09063157i&0.45796188-0.08837374i
\\
 \hline \hline
\end{tabular}
\caption{The fundamental ($n=0$) quasinormal frequencies of  electromagnetic field for $l=1 $ in the four-dimensional scalar hairy black hole spacetime (\ref{m1}).}
\label{tab1}
\end{table}
\begin{table}[ht]
\center
\begin{tabular}[b]{cccc}
 \hline \hline
 \;\;\;\; $c$ \;\;\;\; & \;\;\;\; $\omega\ \ \ (\alpha=0.3)$\;\;\;\;  & \;\;\;\;  $\omega \ \ \ (\alpha=0.4)$\;\;\;\;& \;\;\;\; $\omega \ \ \ (\alpha=0.5)$ \;\;\;\; \\ \hline
\\
2.0&\; 0.53789806-0.09533202i \;&\; 0.59287909-0.09537478i \;&\; 0.65710166-0.09501110i \;
 \\
4.0&0.53789262-0.09533242i&0.59285084-0.09526655i&0.65728145-0.09439396i
 \\
6.0&0.53788718-0.09533282i&0.59282131-0.09515920i&0.65743594-0.09376814i
 \\
8.0&0.53788174-0.09533322i&0.59279055-0.09505276i&0.65756390-0.09313661i
\\
10 &0.53787631-0.09533362i&0.59275859-0.09494727i&0.65766504-0.09250269i
\\
12 &0.53787088-0.09533402i&0.59272548-0.09484273i&0.65774019-0.09186982i
\\
\hline\hline
$c$  &$\omega \ \ \ (\alpha=0.55)$ & $\omega \ \ \ (\alpha=0.60)$ & $\omega \ \ \ (\alpha=0.70)$\\ \hline
 \\
0.1&0.69153046-0.09562530i&0.72757470-0.09565475i&0.80316608-0.09569537i
 \\
0.5&0.69165014-0.09542586i&0.72782053-0.09536380i&0.80381498-0.09520926i
 \\
1.0&0.69179764-0.09517359i&0.72812558-0.09499245i&0.80463046-0.09457567i
 \\
2.0&0.69208496-0.09465924i&0.72872630-0.09422384i&0.80626998-0.09321518i
\\
3.0&0.69236083-0.09413221i&0.72931088-0.09342039i&0.80750659-0.09171888i
\\
4.0&0.69262384-0.09359312i&0.72987467-0.09258249i&0.80951762-0.09007625i
\\ \hline\hline
\end{tabular}
\caption{The fundamental ($n=0$) quasinormal frequencies of  electromagnetic field for $l=2 $ in the four-dimensional scalar hairy black hole spacetime (\ref{m1}).}
\label{tab2}
\end{table}
\begin{figure}[ht]
\center
\includegraphics[width=6cm]{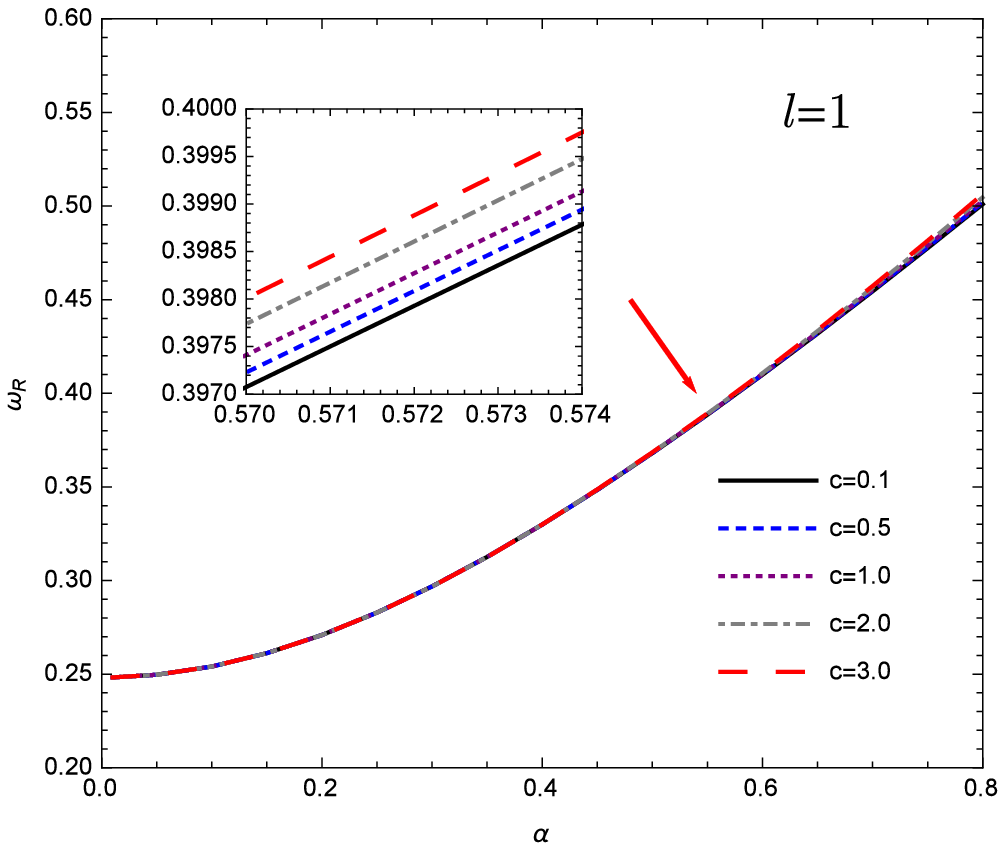}\includegraphics[width=6cm]{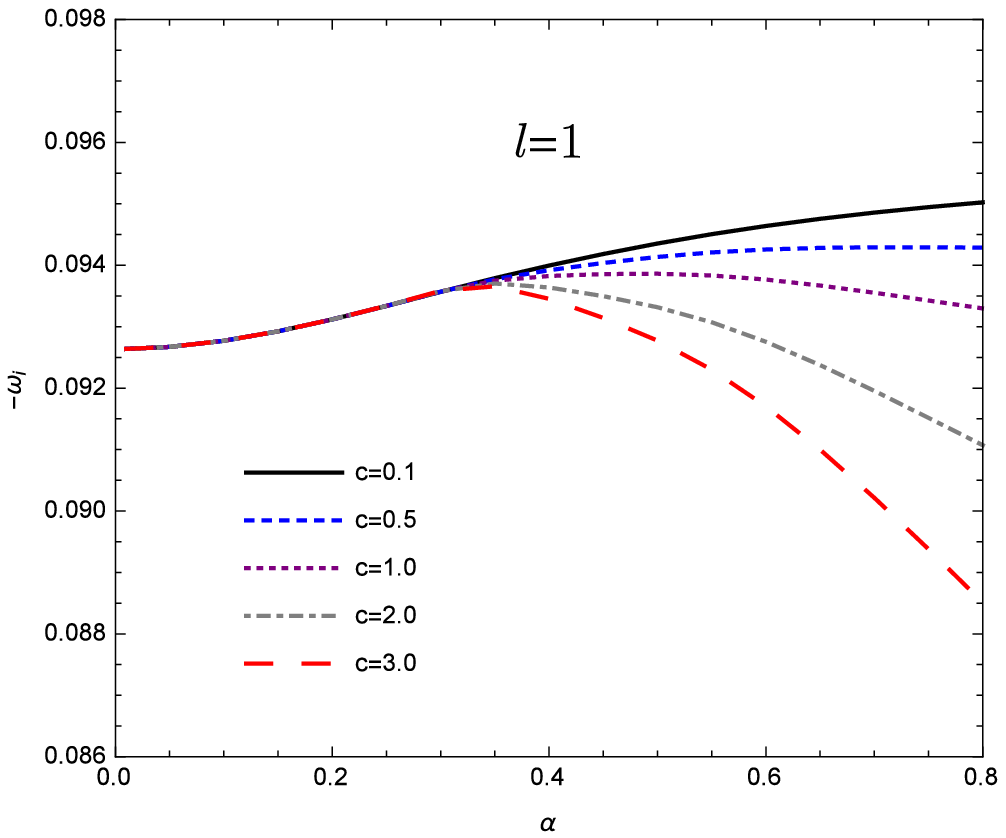} \\
\includegraphics[width=6cm]{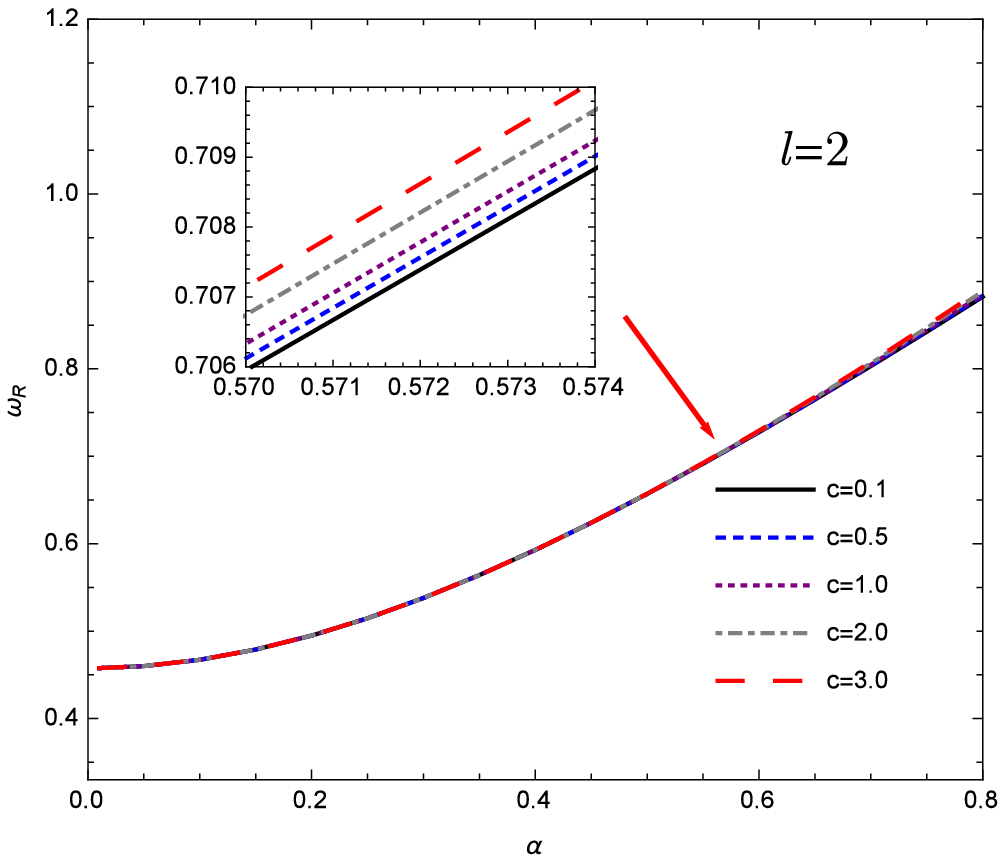}\includegraphics[width=6.2cm]{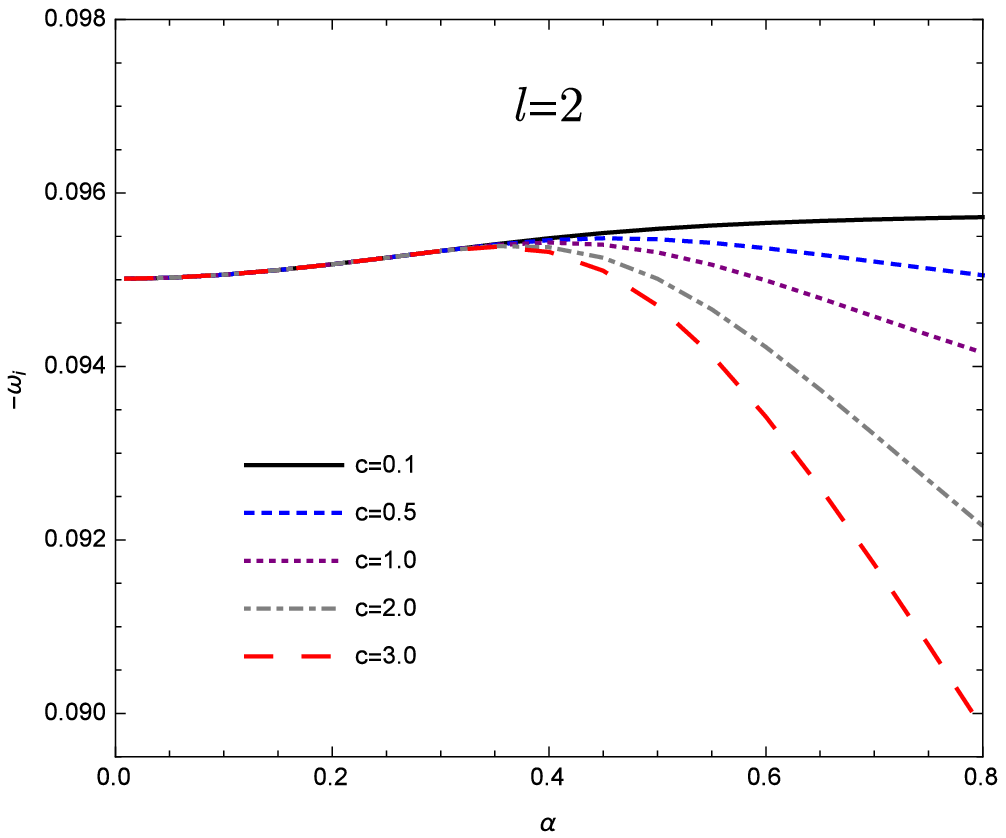}
\caption{Change of  quasinormal
frequencies ($n=0$) with the parameter $\alpha$ for electromagnetic field perturbations with fixed $c$. The upper and bottom rows correspond to the cases $l=1$ and $l=2$, respectively. }
\label{fig2}
\end{figure}
\begin{figure}[ht]
\center
\includegraphics[width=6cm]{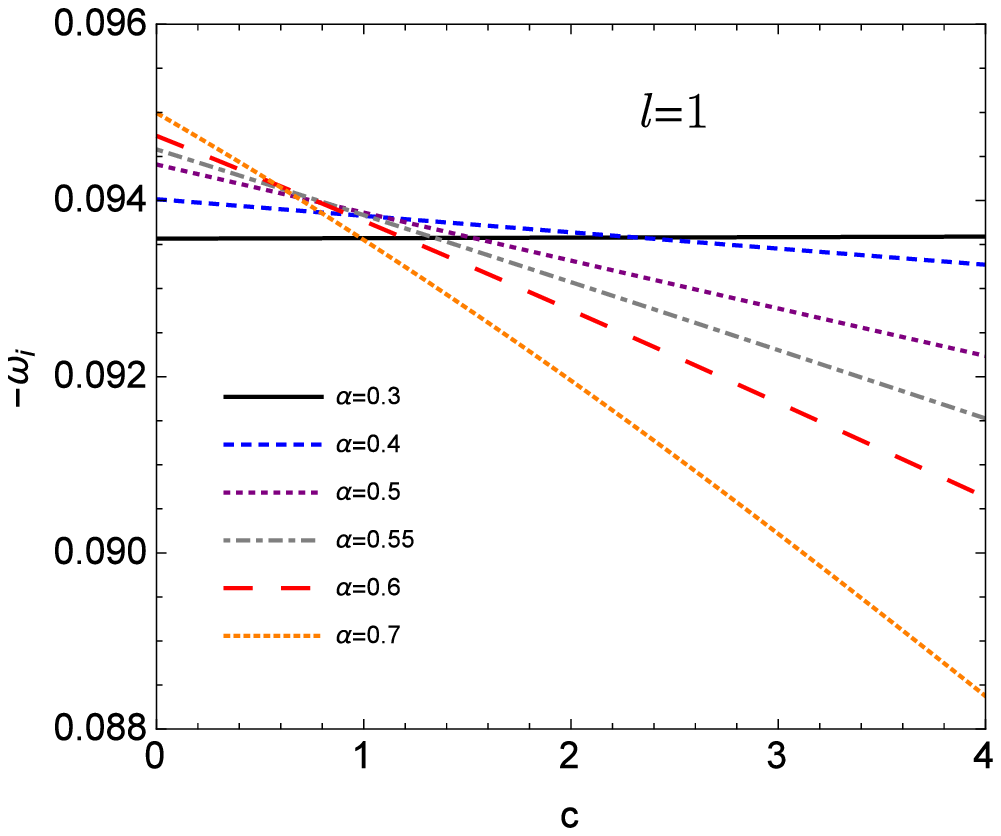}\includegraphics[width=6cm]{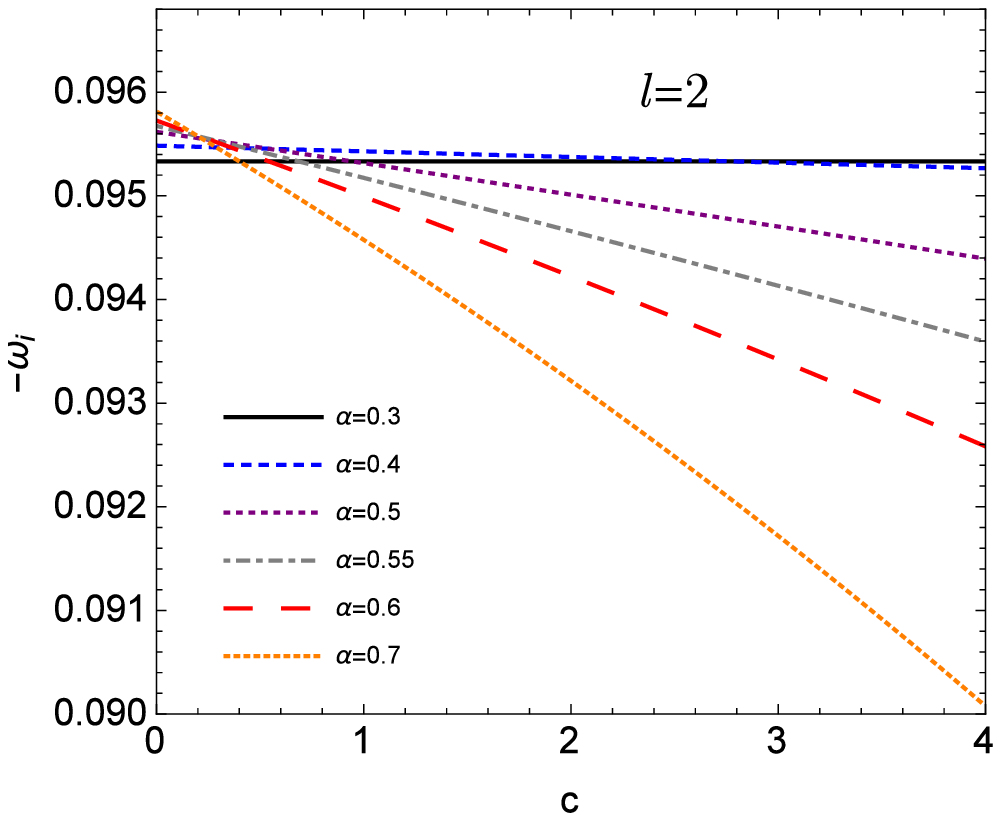}
\caption{Dependence of the  imaginary part of quasinormal
frequencies ($n=0$) on the parameter $c$ for electromagnetic field perturbations with fixed $\alpha$. The left and right planes correspond to the cases $l=1$ and $l=2$, respectively. }
\label{fig3}
\end{figure}
\begin{figure}[ht]
\center
\includegraphics[width=5.3cm]{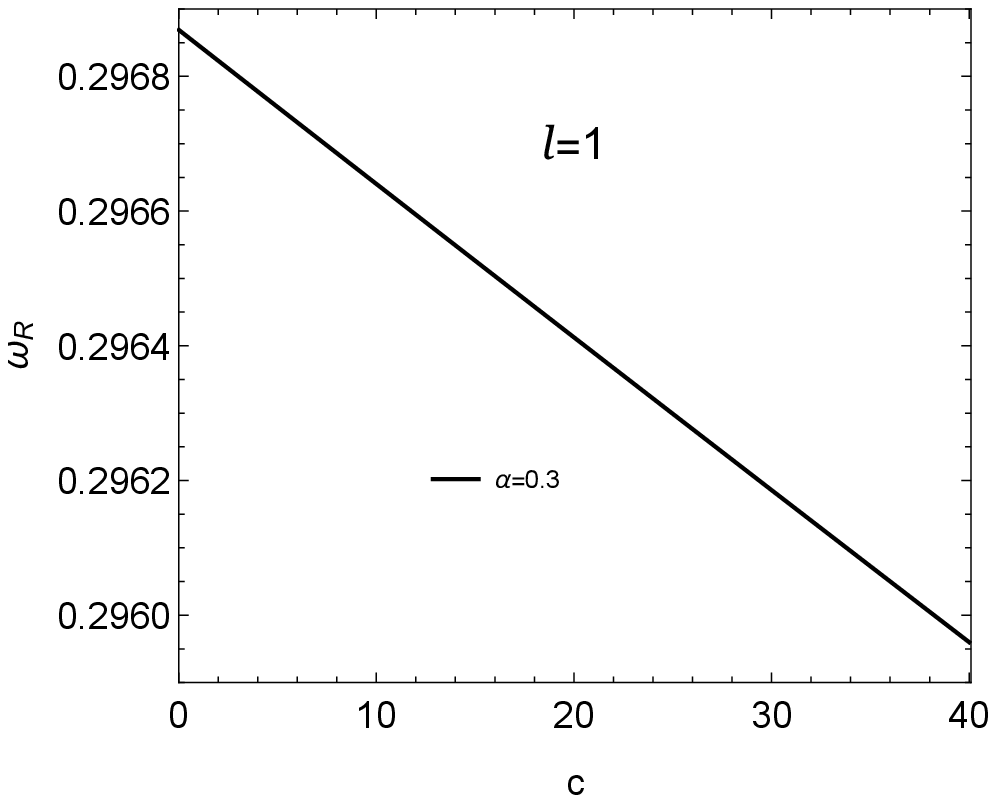}\includegraphics[width=5.2cm]{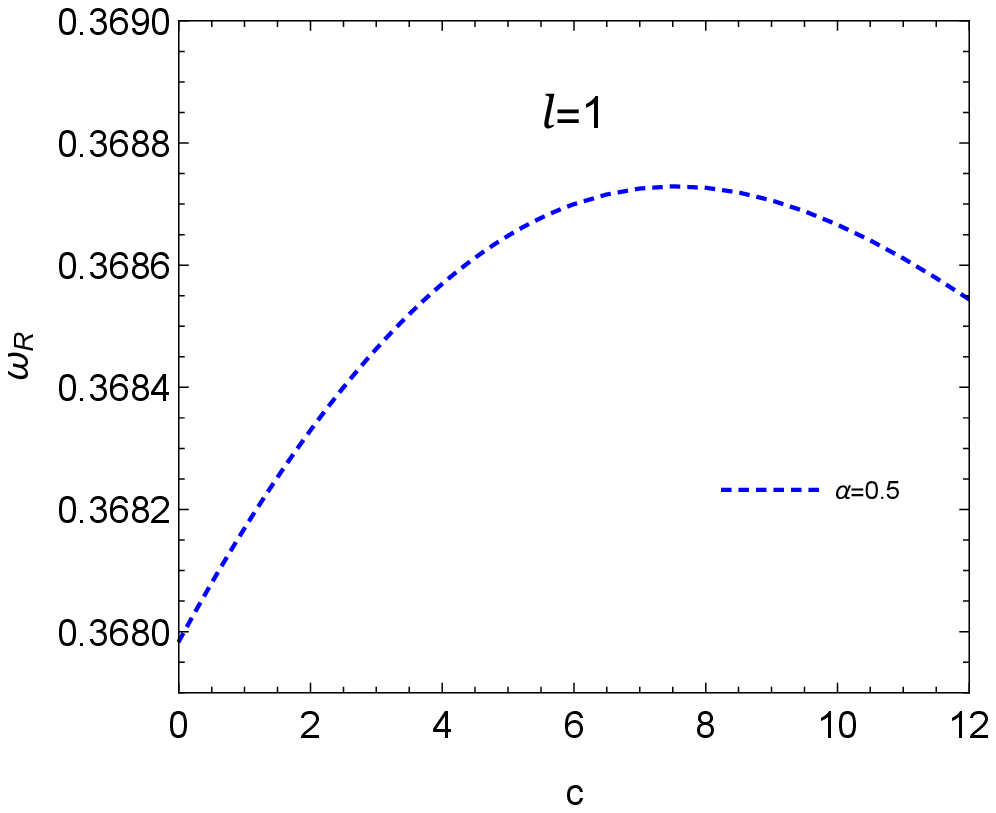} \includegraphics[width=5cm]{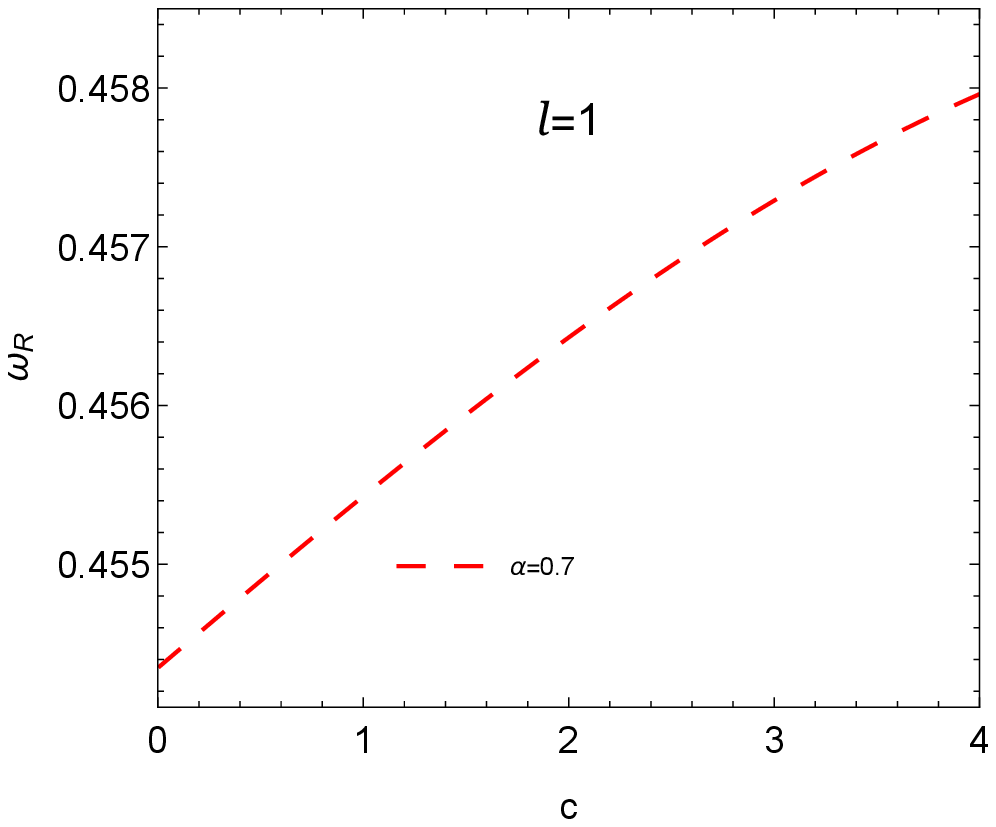}\\
\includegraphics[width=5.4cm]{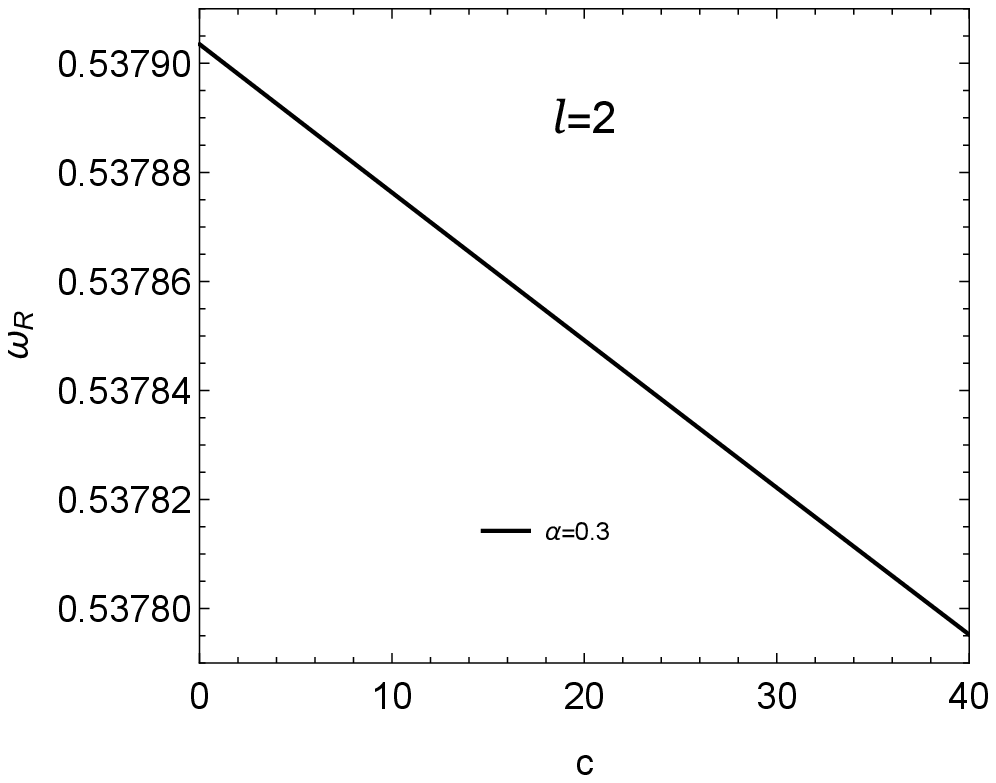}\includegraphics[width=5.2cm]{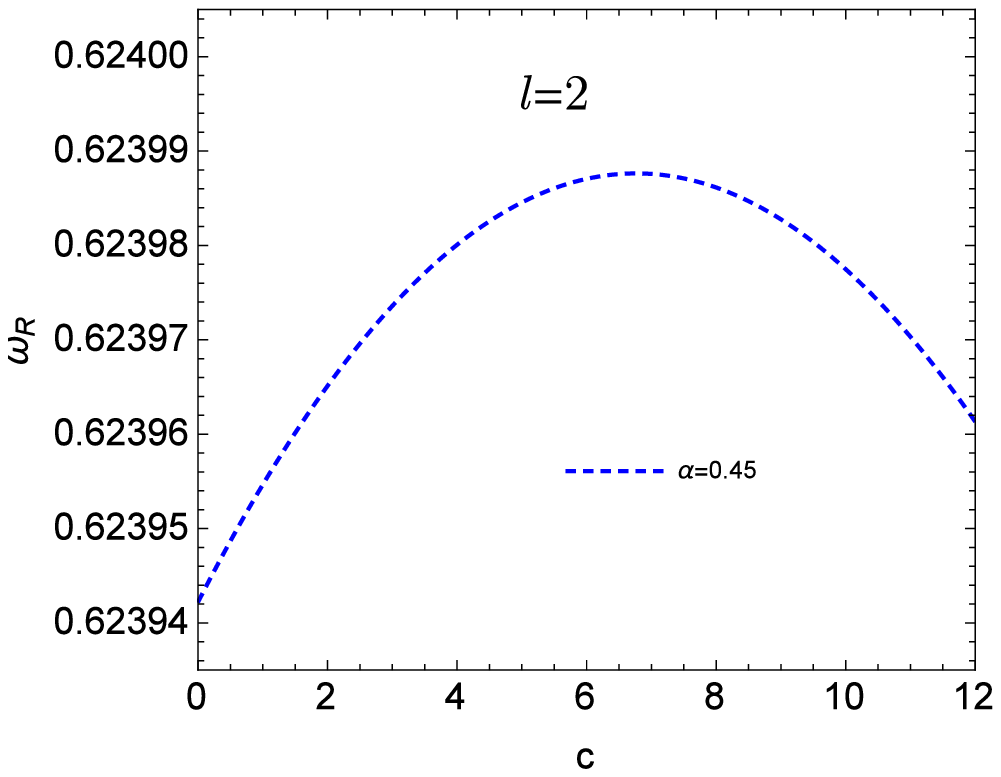} \includegraphics[width=5cm]{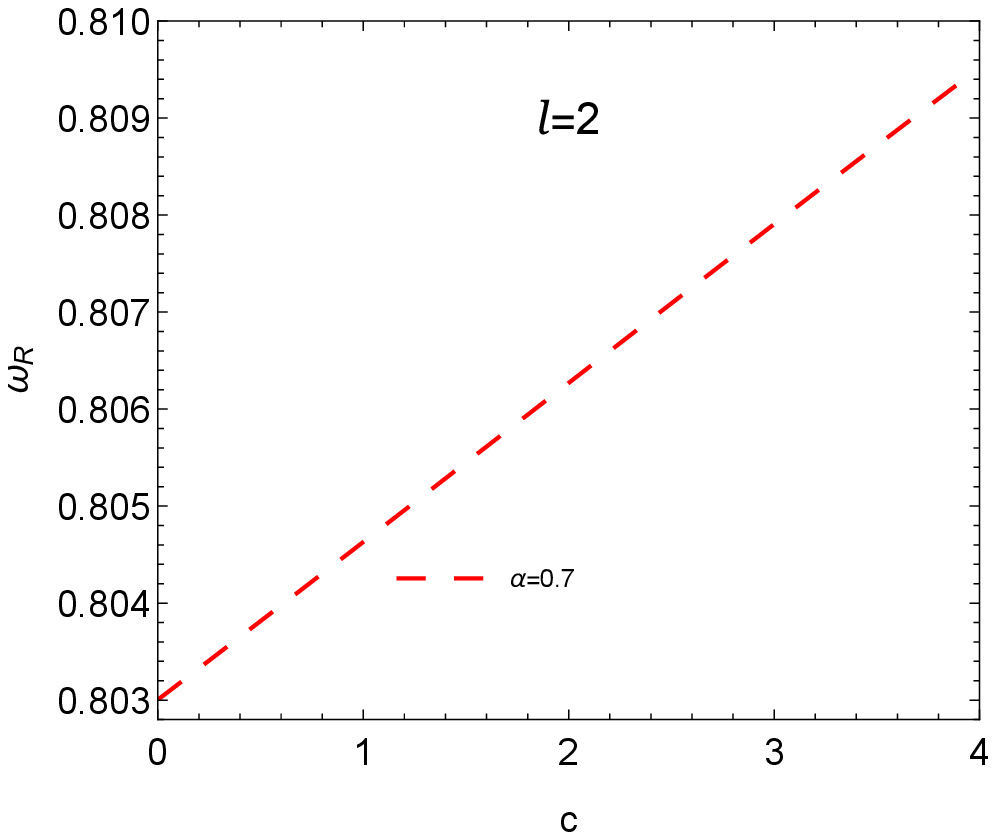}
\caption{Dependence of the  real part of quasinormal
frequencies ($n=0$) on the parameter $c$ for electromagnetic field perturbations with fixed $\alpha$. The upper and bottom rows correspond to the cases $l=1$ and $l=2$, respectively. }
\label{fig4}
\end{figure}

We here list the fundamental quasinormal
frequencies for $l=1$ and $l=2$ in Tables \ref{tab1} and \ref{tab2}, respectively. It is obvious that quasinormal frequencies depend on parameters $\alpha$ and $c$. For fixed $c$,  we find from Tables \ref{tab1} and \ref{tab2} that  the real part of quasinormal frequencies increase monotonously with $\alpha$ for $l=1$ and $l=2$, which is shown in Fig.(\ref{fig2}). However, the change of the absolute value of imaginary part with $\alpha$ becomes more complicated.  With the increase of $\alpha$, from Fig.(\ref{fig2}), we find that the absolute value of imaginary part increases monotonously for the cases with the smaller $c$, but it first increases and then decreases for the cases with the larger $c$.
Figs.(\ref{fig3})-(\ref{fig4}) tell us that for fixed $\alpha$, the change of the absolute value of imaginary part with the parameter $c$ becomes more simple and they are the monotonously decreasing functions of $c$ in both cases $l=1$ and $l=2$, but the real parts become more complex. The
real parts for both $l=1$ and $l=2$ decrease  monotonously with the parameter $c$ for the smaller $\alpha$. With the increase of $\alpha$, the real parts become gradually a first increasing and then decreasing function of $c$,  and finally they become a strictly increasing function of $c$.  Moreover, from Figs.(\ref{fig2})-(\ref{fig3}), we can obtain that the dependence of quasinormal frequencies on $c$ is no significant as $\alpha$ is smaller, which is understandable since the contribution of term containing $c$ to the effect potential $V(r)$ in Eq. (\ref{radial}) is tiny in the small $\alpha$ cases.

Finally, we extend our discussion to the late-time tail behavior
of the electromagnetic perturbation in the background of a scalar hairy black hole spacetime (\ref{m1}).
It is well known that the late-time behavior of perturbation fields is determined by the backscattering
from asymptotically far regions and the leading contribution to
the Green's function comes from the low-frequency parts. Adopting the low-frequency approximation and neglecting terms of order $O( (\frac{M}{r})^2 )$, one can  expend the
wave equation (\ref{radial}) for the electromagnetic field as a power
series in $M/r$
\begin{eqnarray}
\left[\frac{d^2}{dr^2}+\omega^2+\frac{4M\omega^2}{r}-\frac{\rho^2-\frac{1}{4}}{r^2}\right]\zeta(r,\omega)=0,
\end{eqnarray}
with
\begin{eqnarray}
\zeta(r,\omega)=\sqrt{1-\frac{2M}{r}+\frac{c}{r^{(4+1/\alpha^2)}}}\Psi,\;\;\;\;\;\;\;\;\;
\rho=\sqrt{(l+\frac{1}{2})^2+4\alpha^2l(l+1)}.
\end{eqnarray}
After some similar operations in Refs.\cite{latt1,latt2,latt3}, we find that in the background of a scalar hairy black hole spacetime (\ref{m1}), the asymptotic behavior of the electromagnetic perturbation at timelike infinity is described by the Green function
\begin{eqnarray}
G^C(r_*,r'_*;t)&=&\frac{2^{2\rho}M[\Gamma{(\rho+\frac{1}{2})}]^2(-1)^{2\rho+1}\Gamma{(2\rho+2)}}{\pi
[\Gamma{(2\rho+1)}]^2}(r'_*r_*)^{\rho+1/2} t^{-2\rho-2}\sim  t^{-2\sqrt{(l+\frac{1}{2})^2+4\alpha^2l(l+1)}-2}.
\end{eqnarray}
Thus, for the electromagnetic perturbation in the background of a scalar hairy black hole spacetime (\ref{m1}), the late-time tails are
dominated not only by the multiple moment $l$, but also by the parameter $\alpha$, which is similar to that in the global monopole spacetime. It can be explained by a fact that these spacetimes own the similar asymptotical structure at spatial infinity due to the existence of solid angle deficit.
As $\alpha\rightarrow 0$,  the late-time tail becomes $t^{-(2l+3)}$, which is consistent with that in usual Schwarzschild black hole spacetime. Moreover, we find that the presence of the parameter $\alpha$ makes the field decay more rapidly. It
implies that the spacetime properties arising from the logarithmic cubic term $G_3(\phi,X)=-\alpha M_P\log(-X)$ in the action (\ref{acts})
 play an important role in the dynamical evolutions of the
electromagnetic perturbation in the background of a scalar hairy black hole (\ref{m1}).

\section{summary}

In this paper, we have investigated dynamical evolution of electromagnetic perturbation in a scalar hairy black hole spacetime, which belongs to solutions in Horndeski theory with a logarithmic cubic term. The parameters $\alpha$ and $c$ affect the existence of event horizon and modify the asymptotical structure of spacetime at spatial infinity due to the existence of solid angle deficit. These properties caused by the spacetime parameters affect sharply the quasinormal frequencies and late-time tail behaviors of electromagnetic perturbation. For fixed $c$, our results show that the real part of quasinormal frequencies increase monotonously with $\alpha$ for $l=1$ and $l=2$. However, the change of the absolute value of imaginary part with $\alpha$ depends on the angular quantum
number $l$, the spacetime parameters $c$ and $\alpha$. With the increase of $\alpha$, the absolute value of imaginary part increases monotonously for the cases with the smaller $c$, but it first increases and then decreases for the cases with the larger $c$. For fixed $\alpha$, the change of the absolute value of imaginary part with the parameter $c$ are the monotonously decreasing functions of $c$ in both cases $l=1$ and $l=2$, but the real part becomes more complicated. The
real parts for both $l=1$ and $l=2$ decrease  monotonously with the parameter $c$ for the smaller $\alpha$. With the increase of $\alpha$, the real parts become gradually a first increasing and then decreasing function of $c$,  and finally they become a strictly increasing function of $c$. Finally, we find that the late-time tails of electromagnetic perturbation
in the background of a scalar hairy black hole spacetime (\ref{m1}) depend also on the parameter $\alpha$ due to the existence of solid angle deficit. The presence of the parameter $\alpha$ makes the field decay more rapidly. It implies that the spacetime properties arising from the logarithmic cubic term $G_3(\phi,X)=-\alpha M_P\log(-X)$ in the action (\ref{acts})
 play an important role in the dynamical evolutions of the
electromagnetic perturbation in the background of a scalar hairy black hole (\ref{m1}).

\section{\bf Acknowledgments}
We would like to thank Profs. A. Vikman  and  R. A. Konoplya for their useful comments.
This work was partially supported by the National Natural Science Foundation of China under
Grant No. 11875026, the Scientific Research
Fund of Hunan Provincial Education Department Grant
No. 17A124. J. Jing's work was partially supported by
the National Natural Science Foundation of China under
Grant No. 11475061, 11875025.

\end{document}